\documentclass[fleqn,12pt,twoside]{article}
\usepackage[headings]{espcrc1}

\readRCS
$Id: espcrc1.tex,v 1.2 2004/02/24 11:22:11 spepping Exp $
\usepackage{graphicx}
\usepackage[figuresright]{rotating}

\newcommand{\AmS}{{\protect\the\textfont2
  A\kern-.1667em\lower.5ex\hbox{M}\kern-.125emS}}

\title{Uncovering the secrets of the 2d random-bond Blume-Capel model}

\author{A. Malakis\address[MCSD]{Department of Physics, Section of Solid
State Physics, University of Athens, Panepistimiopolis, GR 15784
Zografos, Athens, Greece}%
        \thanks{amalakis@phys.uoa.gr},
        A. Nihat Berker\address{Faculty of Engineering and Natural Sciences, Sabanci University, Orhanli, Tuzla
        34956, Istanbul, Turkey}\address{Feza
G\"{u}rsey Research Institute, T\"{U}B\.{I}TAK - Bosphorus
University, \c{C}engelk\"{o}y 34684, Istanbul,
Turkey}\address{Department of Physics, Massachusetts Institute of
Technology, Cambridge, Massachusetts 02139, U.S.A.},
        I. A. Hadjiagapiou\addressmark[MCSD],
        N. G. Fytas\addressmark[MCSD]\
        and
        T. Papakonstantinou\addressmark[MCSD]}

\runtitle{Uncovering the secrets of the 2d random-bond Blume-Capel
model} \runauthor{A. Malakis}

\begin{document}

\maketitle

\begin{abstract}
The effects of bond randomness on the ground-state structure,
phase diagram and critical behavior of the square lattice
ferromagnetic Blume-Capel (BC) model are discussed. The
calculation of ground states at strong disorder and large values
of the crystal field is carried out by mapping the system onto a
network and we search for a minimum cut by a maximum flow method.
In finite temperatures the system is studied by an efficient
two-stage Wang-Landau (WL) method for several values of the
crystal field, including both the first- and second-order phase
transition regimes of the pure model. We attempt to explain the
enhancement of ferromagnetic order and we discuss the critical
behavior of the random-bond model. Our results provide evidence
for a strong violation of universality along the second-order
phase transition line of the random-bond version.
\end{abstract}

\section{INTRODUCTION}
\label{sec:1}

Although originally thought to play a rather innocuous role,
quenched bond randomness may (or may not) modify the critical
exponents of second-order phase transitions~\cite{harris74},
whereas in 2d it always affects first-order phase transitions by
conversion to second-order phase transitions even for
infinitesimal strength~\cite{aizenman89,hui89,berker93}. These
predictions have been confirmed by various Monte Carlo simulations
and they have been also well verified by our recent numerical
studies via a two-stage WL method~\cite{wang01,fytas08,malakis09}.
The random-bond version of the BC model is defined by the
Hamiltonian
\begin{equation}
\label{eq:1} H=-\sum_{<ij>}J_{ij}s_{i}s_{j}+\Delta
\sum_{i}s_{i}^{2}\;;\
P(J_{ij})=\frac{1}{2}~[\delta(J_{ij}-J_{1})+\delta(J_{ij}-J_{2})]\;\;;\;r=J_{2}/J_{1},
\end{equation}
where the spin variables $s_{i}$ take on the values $-1, 0$, or
$+1$, $<ij>$ indicates summation over all nearest-neighbor pairs
of sites and $\Delta$ is the value of the crystal field. The above
bimodal distribution describes our choice for the quenched random
interactions corresponding to disorder strength $r=J_{2}/J_{1}$,
where both $J_{1}$ and $J_{2}$ are taken positive and the
temperature scale may be set by fixing $2k_{B}/(J_{1}+J_{2})=1$.
The pure square lattice model exhibits a phase diagram with
ordered ferromagnetic and disordered paramagnetic phases separated
by a transition line that changes from an Ising-like phase
transition to a first-order transition at a tricritical point
$(T_{t},\Delta_{t})=(0.609(4),1.965(5))$
~\cite{blume66,beale86,landau86,silva06}.

The present paper considers further interesting aspects of the
random-bond version of the 2d BC model. In particular we will
report on the effects of bond randomness on the ground-state
structure, the phase diagram and the critical behavior of the
square lattice ferromagnetic BC model. The ground-state problem is
briefly discussed in the next Section and the order-parameter
behavior, as a function of the disorder strength for several
values of the crystal field, is illustrated. Then, in
Section~\ref{sec:3}, we report on the effects of bond randomness
on the phase diagram of the model and the enhancement of
ferromagnetic order is clarified. We continue with a study of the
conversion to a second-order phase transition, due to the
introduction of bond-randomness of the first-order transition of
the pure model at the value $\Delta=1.975$. The more general
critical behavior of the second-order phase transition of the
random-bond 2d BC model is also briefly discussed. The implemented
two-stage method of a restricted entropic WL sampling has been
described in our recent studies~\cite{fytas08,malakis09}. Finally,
our conclusions are summarized in Section~\ref{sec:4}.

\section{UNSATURATED FERROMAGNETIC GROUND STATES}
\label{sec:2}

In the presence of bond randomness the competition between the
ferromagnetic interactions with the crystal field may result in
destabilization of the ferromagnetic ground state. This phenomenon
occurs for strong disorder and sufficiently large values of the
crystal field giving rise to unsaturated ground states. To clarify
this possibility, we observe that there will be on the average (in
a large sample of disorder realizations) a finite portion of about
$6.25\%$ of lattice sites with all their connections being weak
couplings ($J_{2}$). Thus, at strong disorder (small $J_{2}$)
these lattice sites will be, at $T=0$, in the $s_{i}=0$ state
provided that $\Delta > 4J_{2}$. Consequently, for strong disorder
and for any $\Delta > 4J_{2}$, vacant sites ($s_{i}=0$) will be
distributed in the ground state of any given disorder realization.
The complexity of this phenomenon may be further illuminated by
defining weak clusters such that, in $T=0$, these weak clusters
will be in the all $s_{i}=0$ state.

This subject has not be studied before and the consequences of the
unsaturated ground state on the critical behavior of the 2d (and
3d) random-bond BC model are not known.
\begin{figure}[htbp]
\includegraphics*[width=8 cm]{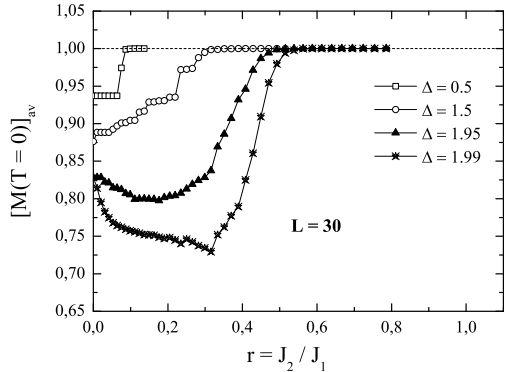}
\caption{\label{fig:1}Ground-state behavior of the order parameter
of the random-bond 2d BC model versus $r$ for various values of
$\Delta$ averaged over $250$ disorder realizations.}
\end{figure}
The calculation of the ground states can be carried out in
polynomially bounded computing time by mapping the system into a
network and searching for a minimum cut by using a maximum flow
algorithm. The results shown in figure~\ref{fig:1}, obtained via
the FORD-FULKERSON method (see for instance~\cite{hartmann02}),
illustrate the ground-state behavior of the order parameter
$[M(T=0)]_{av}$ as a function of the disorder strength $r$, for
several values of the crystal field $\Delta=0.5, 1.5, 1.95$, and
$1.99$ using a square lattice of linear size $L=30$. The growing
importance of the $s_{i}=0$ state as we increase the value of the
crystal field reflects the weak clusters complexity mentioned
above and it will be very interesting to find any possible
interconnections with the critical properties of the random-bond
model.

\section{ENHANCEMENT OF FERROMAGNETIC ORDER AND CRITICAL BEHAVIOR}
\label{sec:3}

In general, the introduction of bond randomness is expected to
decrease the phase-transition temperatures which at the
percolation limit of randomness ($r=0$ and $J_{2}=0$) should tend
to zero. However, for weak randomness only a slight decrease is
expected, if the average bond strength is maintained, as
implemented here. Such a slight decrease in the critical
temperature (by $1\%$) was reported earlier, for $\Delta=1$,
whereas, in sharp contrast and for the same disorder strength
($r=0.6$), for $\Delta=1.975$ (first-order regime of the pure
model) we have found a considerable increase of the critical
temperature, by $9\%$~\cite{malakis09}. A microscopic explanation
of this phenomenon was attempted in~\cite{malakis09} by discussing
the behavior of the connectivity spin densities, $Q_{n}=\langle
s_{i}^{2}\rangle_{n}$, where the subscript $n=0,1,2,3,4$ denotes
the class of lattice sites and is the number of the quenched
strong couplings ($J_{1}$) connecting to each site in this class.
\begin{figure}[htbp]
\includegraphics*[width=16 cm]{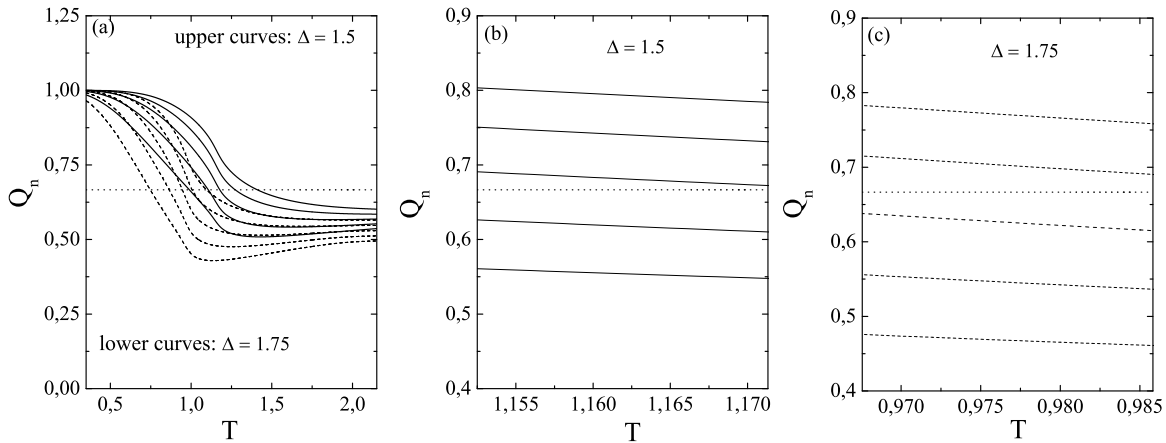}
\caption{\label{fig:2}Average ($10$ realizations) behavior of the
connectivity spin densities $Q_{n}=\langle s_{i}^{2}\rangle_{n}$
for $\Delta=1.5$ and for $\Delta=1.75$. The lattice linear size
used is $L=40$.}
\end{figure}
For $\Delta=1.975$, a pronounced preference was observed for the
$s_{i}=0$ state on the low strong-coupling connectivity sites
whereas the $s_{i}=\pm 1$ states preferentially occurred with
strong-coupling connectivity. It was then pointed out that this
naturally leads to a higher transition temperature and effectively
carries the ordering system to higher non-zero spin densities, the
domain of second-order phase transitions. A similar
microsegregation phenomenon, of the $s_{i}=\pm 1$ states and of
the $s_{i}=0$ state, has been seen in the low-temperature
second-order transition between different ordered phases under
quenched randomness~\cite{kaplan08}.

As verified by our subsequent studies (for $r=0.6$) this
enhancement of the ferromagnetic order takes place in the
neighborhood of $\Delta=1.64$, where the phase diagrams of the
pure and the random-bond models cross each other.
Figure~\ref{fig:2}(a) illustrates the temperature behavior of the
connectivity densities. The illustration gives the behavior for
two values of $\Delta$, namely $\Delta=1.5$ and $\Delta=1.75$
which lie below and above the estimated phase diagrams crossing.
Furthermore, figures~\ref{fig:2}(b) and (c) illustrate the heights
and the the differences (between the smallest $n=0$ and the
largest $n=4$) of the connectivity densities in the corresponding
pseudocritical region, defined by the specific heat and magnetic
susceptibility finite-size peaks.
\begin{figure}[htbp]
\includegraphics*[width=16 cm]{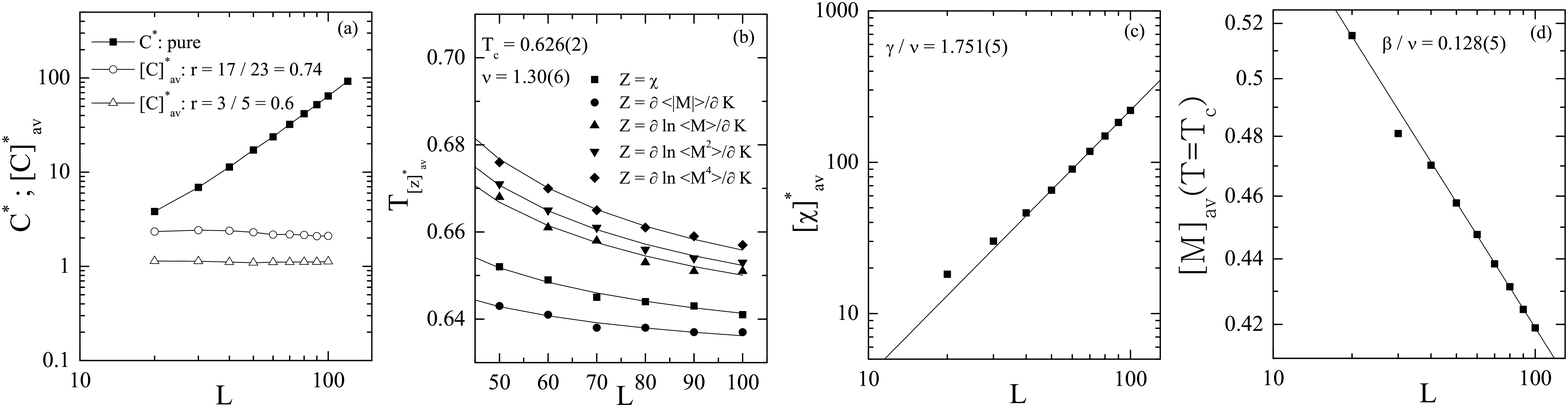}
\caption{\label{fig:3}Behavior of the 2d BC model at
$\Delta=1.975$: (a) Illustration of the saturation of the specific
heat for the random-bond model (open symbols). FSS behavior, for
$r=0.6$, of (b) pseudocritical temperatures, (c) susceptibility
peaks, and (d) order parameter at $T_c$. Linear fits are applied
for $L\geq 50$.}
\end{figure}
For $\Delta=1.75$, the difference is roughly $0.32$, whereas for
$\Delta=1.5$ the difference between the smallest and the largest
of the connectivity densities is only $0.239$. The middle curves
correspond to the leading connectivity density defined on the
lattice sites with two strong and two weak couplings ($n=2$).
These sites amount on the average to a relative majority of
$37.5\%$ of the lattice sites. The height of the middle curve
falls below the value $2/3$ (shown by the dot-line parallel to
$T$-axis) as we move from the value $\Delta=1.5$ to the value
$\Delta=1.75$ signaling that the preference on the $s_{i}=0$ state
is now extended even to the relative majority connectivity sites.
This observation may be related to the enhancement of the
ferromagnetic order as we pass through the crossing point of the
two phase diagrams at $\Delta=1.64$. Thus, in agreement with our
previous conclusion microsegregation does occur in the first-order
regime of the pure model where macrosegregation occurs in the
absence of bond randomness. However, the enhancement of
ferromagnetic order has already appeared before the first-order
regime of the pure model as a kind of a precursor effect induced
by weak randomness.

We now consider the critical behavior at
$\Delta=1.975$~\cite{malakis09}. Figure~\ref{fig:3}(a) contrasts
the specific heat results for the pure 2d BC model and two
strengths of disorder, $r=17/23\simeq 0.74$ and $r=3/5=0.6$. The
saturation of the specific heat is clear and signals the
conversion of the first-order transition to a second-order
transition with a negative critical exponent $\alpha$.
Figures~\ref{fig:3}(b) - (d) summarize our finite-size scaling
(FSS) analysis for the disorder strength $r=3/5=0.6$. The behavior
of five pseudocritical temperatures $T_{[Z]^{\ast}_{av}}$,
$T_{[Z]^{\ast}_{av}}=T_{c}+bL^{-1/\nu}$ (corresponding to the
peaks of susceptibility, derivative of the absolute order
parameter with respect to the inverse temperature ($K=1/T$) and
first-, second-, and fourth-order logarithmic derivatives of the
order parameter with respect to the inverse temperature) is
presented in figure~\ref{fig:3}(b) and allow us to estimate
$T_{c}=0.626(2)$ and for the correlation length exponent
$\nu=1.30(6)$. From the peaks of the average susceptibility and
the values of the average order parameter at $T_{c}=0.626$, we
have estimated, as shown in Figures~\ref{fig:3}(c) and (d), that
the values of the exponent rations $\gamma/\nu=1.751$ and
$\beta/\nu=0.128$ are very close to the exact values of the Ising
model. This ``weak universality'' is well obeyed for several pure
and disordered models, including the pure and random-bond version
of the square Ising model with nearest- and next-nearest-neighbor
competing interactions~\cite{fytas08}.

At $\Delta=1$ the random version should be comparable with the
random Ising model, a model that has been extensively investigated
and
debated~\cite{fytas08,dotsenko81,shalaev84,shankar87,ludwig87,reis96,ballesteros97,selke98}.
Fitting our data for the corresponding pseudocritical temperatures
in the range $L=50-100$ to the expected power-law behavior
mentioned above, we find that the critical temperature is
$T_{c}=1.3812(4)$ and the shift exponent is $1/\nu=1.011(22)$.
This last estimate is a strong indication that the random-bond 2d
BC at $\Delta=1$ and weak disorder has the same value of the
correlation's length critical exponent as the 2d Ising model.
Furthermore, our data for the specific heat maxima averaged over
disorder, $[C]_{av}^{\ast}$, showed that the expected
double-logarithmic divergence scenario is well obeyed. Finally,
the FSS behavior of the susceptibility peaks (giving the estimate
1.749(7) for $\gamma/\nu$) and the order-parameter values at the
estimated critical temperature $T_{c}=1.3812$ (giving an estimate
$\beta/\nu=0.126(4)$) are in good agreement with the expected 2d
Ising universality class behavior.

\section{CONCLUSIONS}
\label{sec:4}

The effects of bond randomness on the ground-state structure of
the 2d BC model have been briefly illustrated by mapping the
system onto a network and searching for a minimum cut by using a
maximum flow algorithm. Furthermore, we clarified some aspects of
the enhancement of ferromagnetic order, due to bond randomness,
and we found that this appears before the first-order regime of
the pure model. Our study at $\Delta=1.975$ shows a conversion of
the first-order transition of the pure model to a second-order
phase transition, giving a distinctive universality class with
$\nu=1.30(6)$ and supporting an extensive but weak universality as
in a wide variety of 2d systems without~\cite{suzuki74} and
with~\cite{fytas08,kim94} quenched disorder. At $\Delta=1$, our
study of the random-bond 2d BC model indicates that for weak
disorder the random system belongs to the same universality class
as the random Ising model and the effect of the bond disorder on
the specific heat is well described by the double logarithmic
scenario. These results amount to a strong violation of
universality since the two second-order phase transitions
mentioned above, with different sets of critical exponents, are
between the same ferromagnetic and paramagnetic phases. A more
complete presentation of the effects of strong disorder on the
ground-state structure and also on the critical behavior of the 2d
BC model will appear in a forthcoming paper.

\section*{ACKNOWLEDGEMENTS}

This research was supported by the Special Account for Research
Grants of the University of Athens under Grant No. 70/4/4071.

\end{document}